\documentclass{desyproc}

\begin{document}
\title{Interferometry in pulsed fields}

\author{{\slshape Babette D\"obrich$^1$ and Holger Gies$^{1,2}$}\\[1ex]
  $^1$Theoretisch-Physikalisches Institut, Friedrich-Schiller-Universit\"at
  Jena \\ 
  $^2$Helmholtz Institute Jena\\
Max-Wien-Platz 1, D-07743 Jena, Germany\\ }
\contribID{dobrich\_babette}

\desyproc{DESY-PROC-2009-05}
\acronym{Patras 2009} 
\doi  

\maketitle

\begin{abstract}
We discuss the particle-physics discovery potential of ground-based
gravitational-wave interferometers. With the use of pulsed magnetic fields,
current and future gravitational-wave interferometers could not only be
utilized to observe phenomena of strong-field QED, but they could also be
applied to sweep the parameter space of particles of the hidden sector.
\end{abstract}

\section{Introduction}
The presence of charged quantum vacuum fluctuations induces self-interactions
of the electromagnetic field~\cite{Heisenberg:1936qt}. In particular, light
passing through a strong external magnetic field is expected to
travel at reduced velocity compared to the propagation through plain
vacuum~\cite{Baier,Dittrich:2000zu}.

As we argue in the following, the combination of ground-based
gravitational-wave interferometers and strong pulsed magnetic fields forms an
instrument which is sensitive enough to demonstrate nonlinearities in the
propagation of light and thereby contribute to the research of
strong-field QED \cite{Dobrich:2009}. At the same time, it facilitates a search for light particles beyond our
current standard model of particle physics.

\section{Alternative goals for gravitational-wave interferometers}

In order to detect gravitational-waves by means of interferometry, two evacuated tubes of equal length $L$ are installed orthogonally with respect to each other. The respective tubes have a mirror installed at their ends and thus form a cavity for a laser beam which is directed through both tubes by means of a beam splitter. An incoming gravitational-wave will 
induce a relative change $\Delta L(t)$ among the lengths of the two arms as a function of time. Alternatively, an \textit{apparent} change of optical path length $L$ can be caused by applying an external magnetic field $B(t)$ over a distance $x$ in one of the interferometer arms, as the light traveling through the magnetic field region will propagate at reduced velocity. Using natural units $\hbar=c=1$, this implies a so-called \textit{strain} in the interferometer
\begin{equation}
h(t) = \frac{\Delta L}{L}(t) = \frac{x}{L} (1-v(t)) \ , \label{eq:strain}
\end{equation}
as first suggested by \cite{Boer:2002zw}, cf. also \cite{Denisov_Zavattini}. 

Since the sensitivity of the interferometer to the strain $h(t)$ is limited by diverse sources of noise, the temporal variation of $h(t)$ should be adapted to the region of highest sensitivity. Generically, gravitational-wave interferometers are most sensitive to variations at frequencies of about $\mathcal{O}(100\mathrm{Hz})$. More precisely, the specific sensitivity of each interferometer can be read off its spectral noise density function $S_h(f)$, see e.g.~\cite{Blair:1991wd}. In conclusion, for the detection of nonlinear light propagation with the help of gravitational-wave interferometers one needs magnetic fields varying at the millisecond scale.

In fact, such pulsed fields are provided by several magnetic field laboratories around the world. Focussing on the ongoing research at the Dresden High-Magnetic-Field-Laboratory (HDL)~\cite{wosnitza}, we consider the specifications of a technically feasible Helmholtz-coil setup with a coil diameter of $x=0.2\mathrm{m}$. The need for a Helmholtz setup arises from the fact that no nonlinearities are induced for light traveling along the direction of the magnetic field lines. By contrast, for light traveling orthogonally to the magnetic field lines, the effect is maximized\footnote{For this reason, also the drop-off in field strength perpendicular to the field lines which is generic for Helmholtz coils must be minimized.}, depending on the beam polarization.

A feasible model for $N$ subsequent field pulses is a damped sinusoidal oscillation:
\begin{equation}
B(t)=B_{0}\sum_{i=0}^{N-1}\theta(t-t_{i})
\sin(2 \pi \nu_{B}(t-t_{i}))\exp(-\gamma(t-t_{i}))\ ,
\label{eq:model_pulse}
\end{equation} 
with pulse frequency $\nu_{B}$ and a damping constant $\gamma$. For the following estimates, we assume $B_{\mathrm{max}}=60\mathrm{T}$ and $B_{\mathrm{min}}=-6\mathrm{T}$  which fixes the amplitude $B_{0}\approx148\mathrm{T}$ and relates the remaining parameters via $\gamma=2 \nu_{B} \ln\left|B_{\mathrm{max}}/B_{\mathrm{min}}\right|$.

A meaningful measure for the visibility of the strain $h(t)$ is the signal-to-noise-ratio (SNR) $d$. Its value is a measure for the likeliness that the strain is induced by the external magnetic field rather than due to random noise fluctuations. Applying a matched filter (or ''Wiener filter'')~\cite{Blair:1991wd}, the square of the SNR is given by

\begin{equation}
d^{2}=2\int_{0}^{\infty}\frac{|\tilde{h}(f)|^{2}}{S_{h}(f)}\mathrm{d}f\
,\quad \tilde{h}(f)=\int_{-\infty}^{\infty}h(t)e^{-2\pi ift}\mathrm{d}t,
\label{eq:SNR_def}
\end{equation}
where $\tilde{h}(f)$ is the Fourier transform of the induced strain.
A lever arm for the enhancement of this observable is provided by the fact that the setup for the field pulse is non-destructive and thus the pulse can be repeated after the magnet system has been re-cooled. Depending on the details of the setup, the re-cooling time of the magnet system is on the order of several minutes.
To good accuracy, $N$ subsequent pulses can enhance the SNR by a factor of $\sqrt{N}$:
\begin{equation}
 d^{2}|_{N}\approx N\ d^{2}|_{1}. \label{eq:SNR_N}
\end{equation}

\section{Discovery potential at GEO600 and advanced LIGO}

\begin{figure}
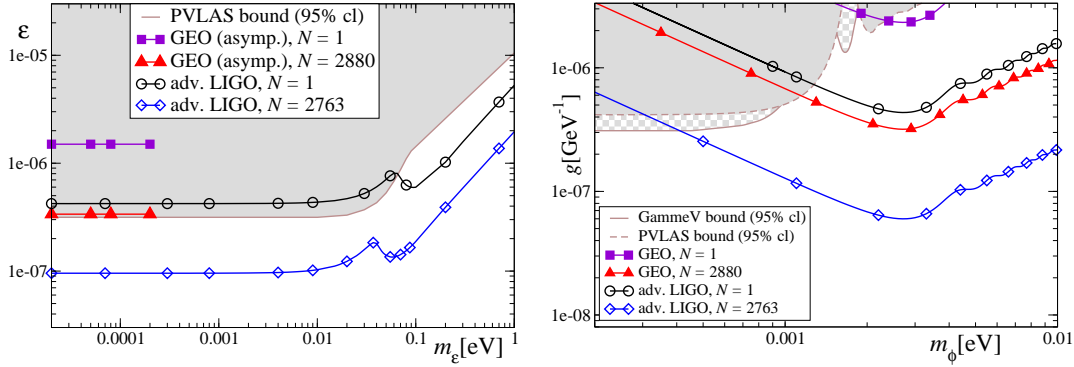

\begin{minipage}{0.49 \linewidth}
\includegraphics[scale=0.285]{doebrich_babette.fig1.eps}
\end{minipage}
\begin{minipage}{0.49 \linewidth}
\includegraphics[scale=0.285]{doebrich_babette.fig2.eps} 
\end{minipage}
\caption{The figure on the left-hand side shows the discovery potential for spin-$\frac{1}{2}$ minicharged particles (MCP), while the figure on the right-hand side applies to axion-like particles (ALP). Already a single pulse measurement at advanced LIGO can improve the best current laboratory bounds~\cite{Zavattini:2005tm,Chou:2007zzc} in the respective coupling-mass planes.} 
\label{fig:figure1}
\end{figure}

We start by computing the number of pulses required to achieve a total SNR of $\mathcal{O}(1)$ for the strain induced by nonlinear QED. To maximize the effect, the laser beam should be polarized in parallel to the external magnetic field lines. The velocity shift then reads~\cite{Baier,Dittrich:2000zu} $1-v=14B^{2}\alpha^{2}/(45 m^{4})$, where $\alpha\approx1/137$ denotes the fine-structure constant and $m$ the electron mass. Together with the parameterization of the field  pulse, see Eq.\eqref{eq:model_pulse}, the velocity shift can be translated into the SNR through Eqs.~\eqref{eq:SNR_def} and~\eqref{eq:strain}, while the number of required pulses $N$ enters through Eq.~\eqref{eq:SNR_N}. We perform the calculation for the noise densities $S_h(f)$ of the advanced LIGO~\cite{ligocurves}, which consists of interferometer arms of length $L=4000\mathrm{m}$, and GEO600~\cite{geocurves}, where $L=600\mathrm{m}$. By a variation of the SNR with respect to the pulse frequency $\nu_B$, we find that for the advanced LIGO $\nu_B \approx 47 \mathrm{Hz}$ yields the greatest strain, while for GEO600 $\nu_B\approx 273 \mathrm{Hz}$ is optimal. In terms of the number of required pulses, this would imply $N \approx 2763$ at advanced LIGO, demanding a continuous operation over a few days, which appears reasonable.
(The operation time at GEO600, however, would be several years since  $N \approx 2.3 \times 10^6$ pulses would be needed for an SNR of $\mathcal{O}(1)$ from the QED induced strain).

In analogy to the vacuum polarization induced by the electron fluctuations, also hypothetical particles with a weak coupling to photons can induce a velocity shift in the interferometer~\cite{Gies:2008wv}. In the following, we therefore deduce the accessible parameter space with respect to coupling and mass for axion-like particles (ALPs) and minicharged particles (MCPs).

The velocity shift induced by fluctuating MCPs~\cite{Gies:2006ca,Ahlers:2006iz} with fractional charge $Q=\epsilon e$ depends strongly on their mass $m_{\epsilon}$. While for large masses, the scaling is analogous to the electromagnetic situation $(1-v)\sim \varepsilon^4 B^2/m_\varepsilon^4$, for low MCP masses the asymptotic limit reads $(1-v)\sim-\varepsilon^{8/3} B^{2/3}/\omega^{4/3}$, where the laser frequency $\omega=1.2 \mathrm{eV}$ for the interferometers.
We consider only MCP masses with a Compton wavelength smaller than the separation of the Helmholtz coils  $\sim\mathcal{O}(1\mathrm{cm})$, implying $m_{\varepsilon}\gtrsim 2\times10^{-5}\mathrm{eV}$. For smaller masses, the homogeneous-field assumption underlying the prediction for the velocity shift is no longer valid.

Uncharged scalar (S) and pseudo-scalar (P) ALPs couple to the $\bot$ and the $\parallel$ mode of the laser beam in the magnetic field, respectively. The corresponding velocity shifts read~\cite{Maiani:1986md} $1-v_\parallel^{\text{P}}=1-v_\bot^{\text{S}} =B^{2}g^{2}/\left[2 m_{\phi}^{2}
\left(1-\sin(2y)/2y \right)\right]$, where $y=xm_{\phi}^{2}/(4\omega)$ with ALP mass $m_{\phi}$ and coupling $g$.


As displayed in Fig.~\ref{fig:figure1}, already a single-pulse measurement at advanced LIGO can improve the currently best laboratory bounds for MCPs~\cite{Zavattini:2005tm,Chou:2007zzc} and ALPs~\cite{Zavattini:2005tm,Ahlers:2006iz} in the upper mass ranges (comparable to results for $\mathcal{O}(10^3)$ pulses at GEO600). Taking $N=2763$ pulses at advanced LIGO, as needed for the QED effect, current laboratory bounds can be improved almost in the entire mass range.

\section{Conclusions}

Pulsed magnetic fields such as provided by the Dresden High-Magnetic-Field-Laboratory can contribute to the research in the strong-field domain of QED for two reasons. Although they have generically a reduced field extent $x$ in comparison to dipole magnets, they can provide for extremely high field strengths $B$.  Since the velocity shifts induced by nonlinear QED, ALPs and the large mass regime of MCPs scale with $x B^2$, the reduced field extent can well be compensated for, see also~\cite{battesti}. Secondly, their pulse frequency can be well matched to the region of highest sensitivity of gravitational-wave interferometers.
For these reasons, combining strong pulsed magnetic fields with the interferometric techniques provided by modern gravitational-wave interferometers can give access to an unexplored parameter regime of strong field QED and at the same time allow to search for particles of a hidden sector.

\section*{Acknowledgments}
B.D. would like to thank the organizers of the 5th Patras Workshop in Durham for the opportunity to contribute to the workshop on the one hand and even more profit from it on the other. The authors acknowledge support from the DFG under GRK1523, SFB/TR18, and Gi328/5-1.
 

\begin{footnotesize}




\begin{thebibliography}{99}
\bibitem{Heisenberg:1936qt}
W.~Heisenberg and H.~Euler,
Z.\ Phys.\ {\bf 98} (1936) 714; 
%
V.~Weisskopf, Kong. Dans. Vid. Selsk. Math-fys. Medd. {\bf XIV}, 166
(1936); 
%
  J.~S.~Schwinger,
  Phys.\ Rev.\  {\bf 82}, 664 (1951).
\bibitem{Baier}
R.~Baier and P.~Breitenlohner, 
Act.~Phys.~Austriaca {\bf 25}, 212 (1967); 
Nuov.~Cim.~B {\bf 47} 117 (1967); %
  S.~L.~Adler,
  Annals Phys.\  {\bf 67}, 599 (1971).

\bibitem{Dittrich:2000zu} W.~Dittrich and H.~Gies,
%
Springer Tracts Mod.\ Phys.\  {\bf 166}, 1 (2000).

\bibitem{Dobrich:2009}
  B.~Dobrich and H.~Gies,
  Europhys.\ Lett.\  {\bf 87}, 21002 (2009)
  [arXiv:0904.0216 [hep-ph]].


%

\bibitem{Gies:2008wv}
  H.~Gies,
  arXiv:0812.0668, to appear in Eur. Phys. J. D, (2009); 
  J.\ Phys.\ A  {\bf 41}, 164039 (2008).

\bibitem{Boer:2002zw}
  D.~Boer and J.~W.~Van Holten,
  arXiv:hep-ph/0204207.


\bibitem{Denisov_Zavattini}
  V.~I.~Denisov, I.~V.~Krivchenkov and N.~V.~Kravtsov,
  Phys.\ Rev.\  D {\bf 69}, 066008 (2004); 
   G.~Zavattini and E.~Calloni,
  Eur.\ Phys.\ J.\  C {\bf 62}, 459 (2009)
  [arXiv:0812.0345 [physics.ins-det]].





\bibitem{wosnitza}
T. Herrmannsd\"orfer, private communication;
J. Wosnitza, {\em et al.}, 
in 2006 IEEE International Conference on Megagauss Magnetic Field Generation 
and Related Topics, G.F. Kiuttu, {\em et al.} (ed.) 
197 (2008) %


\bibitem{Blair:1991wd}
  D.~G.~Blair,
  {\it The Detection Of Gravitational Waves,}
  Cambridge UP (1991);   B.~F.~Schutz,
  in {\it Proc. Les Houches School on Astrophysical Sources of Gravitational
  Radiation}, Cambridge UP (1995).

\bibitem{ligocurves}
D.~ Shoemaker, private communication;   J.~R.~Smith  [LIGO Scientific Collaboration],
  Class.\ Quant.\ Grav.\  {\bf 26}, 114013 (2009)
  [arXiv:0902.0381 [gr-qc]].

\bibitem{geocurves}
http://www.geo600.uni-hannover.de/geocurves/

\bibitem{Maiani:1986md}
  L.~Maiani, R.~Petronzio and E.~Zavattini,
  Phys.\ Lett.\ B {\bf 175}, 359 (1986); 
%
  G.~Raffelt and L.~Stodolsky,
  Phys.\ Rev.\ D {\bf 37}, 1237 (1988).

\bibitem{Gies:2006ca}
  H.~Gies, J.~Jaeckel and A.~Ringwald,
  Phys.\ Rev.\ Lett.\  {\bf 97}, 140402 (2006).


\bibitem{Zavattini:2005tm}
E.~Zavattini {\it et al.}
,
Phys.\ Rev.\ Lett.\  {\bf 96} (2006) 110406; %
  Phys.\ Rev.\  D {\bf 77}, 032006 (2008).

\bibitem{Ahlers:2006iz}
  M.~Ahlers, H.~Gies, J.~Jaeckel and A.~Ringwald,
  Phys.\ Rev.\  D {\bf 75}, 035011 (2007).

\bibitem{Chou:2007zzc}
  A.~S.~Chou {\it et al.}
,
  Phys.\ Rev.\ Lett.\  {\bf 100}, 080402 (2008).

\bibitem{battesti}
R.~Battesti {\it et al. ,
  Eur.\ Phys.\ J. D {\bf 46}, 323 (2008).}

\end{thebibliography}
%

\end{footnotesize}


\end{document}